# SCALED DOWN EXPERIMENTS FOR A STELLARATOR TYPE MAGNETOSTATIC STORAGE RING

N. Joshi[#], M. Droba, O. Meusel, H. Niebuhr and U. Ratzinger
Institut fuer Angewandte Physik, Goethe University, Frankfurt am Main, Germany.

*Abstract*

The beam transport experiments in toroidal magnets were first described in EPAC08 [1] within the framework of a proposed low energy ion storage ring at Frankfurt University. The experiments with two room temperature 30 degree toroids are needed to design the accumulator ring with closed longitudinal magnetic field levels up to 6-8 T. The test setup aims on developing a ring injection system. The primary beam line for the experiments was installed and successfully commissioned in 2009. A special probe for ion beam detection was installed. This modular technique allows online diagnostics of the ion beam along the beam path. In this paper, we present new results on beam transport experiments and discuss transport and transverse beam injection properties of that system.

## INTRODUCTION

The stellarator type storage ring for accumulation of high current low energy beams up to the MeV range is under design. This ring comprises of curved sectors with longitudinal magnetic field to form the Figure-8 geometry (Fig. 1). The continual longitudinal magnetic field should provide focusing and guiding forces. High magnetic field produced by superconducting coils is desired to minimize drift forces and to achieve high current densities. The room temperature experiments scaled down to 0.6 T were planned to investigate the drift dynamics, space charge effects, beam detection and multiturn beam injection system. The measurements from these experiments are compared with simulations.

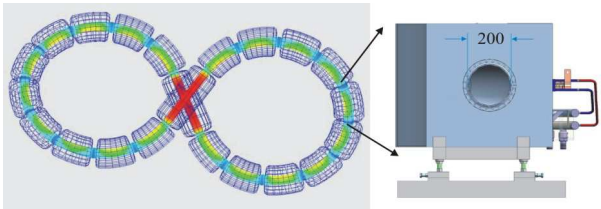

Figure 1: An example of ring formed with toroidal segments depicting single magnetic surface. Magnetic field strength is colour coded. Room temperature segment is shown on the right.

In curved magnetic fields the ion beam shows the Larmor gyration and drift motion, namely the **R** x **B** drift due to curved field lines, grad B drift due to an inhomogeneous field and the **E** x **B** drift due to the space charge [2].

The curvature drift is given by,

$$v_{R \times B} = \frac{mv_\parallel^2}{qB^2} \frac{\vec{R} \times \vec{B}}{R^2} \quad (1)$$

This equation implies ions injected into the magnetic field will experience a vertical drift force in direction perpendicular to the plane formed by the curvature vector **R** and the magnetic field **B**. For a proton beam with energy 10 keV in the magnetic field of 0.6 T and R=1.3 m, it is $v_{RxB}$ = 2.5E4 m/s. Along with the time of flight one can estimate, this drift will be 17 mm along the beam path over arc distance of 680 mm. This drift is directly proportional to the square root of ion mass and inversely proportional to field strength.

The grad B drift arises due to the difference of coil density on the inner side and the outer side of the toroid. The drift velocity due to this inhomogeneity of field is given by,

$$v_{\nabla B} = \frac{1}{2} v_\perp r_L \frac{\vec{B} \times \nabla |B|}{B^2} \quad (2)$$

In the case of the used toroidal segments the magnetic field gradient is 0.4 T/m. For a proton beam with transverse velocity spread up to 1.66E5 m/s, the drift velocity can be maximum $v_{grad\,B}$ =320 m/s.

An additional drift arises due to the self electric field of the beam. The particles will experience additional drift velocity given by,

$$v_{E \times B} = \frac{\vec{E} \times \vec{B}}{B^2} \quad (3)$$

The homogeneous ion beam has a maximum Coulomb force on the boundary surface. Thus the cross product of radial component of the electric field ($E_r$) and the longitudinal magnetic field ($B_\zeta$) gives rise to a rotation of the beam around its own axis. For a proton beam (r = 15 mm) with current I = 3.0 mA at 10 keV energy this drift velocity $v_{E \times B}$ = 4.34E3 m/s.

Thus we conclude in our case the curvature drift is the most dominating over the others.

---

[#]Corresponding author; Email: joshi@iap.uni-frankfurt.de

# EXPERIMENTS

## Beam transport in single toroidal segment

In the first stage the experimental setup consisted of a single sector magnet. The volume type ion source was used. A solenoid matches the beam into the sector magnet with toroidal geometry. Typical operational parameters for individual components are listed in Table 1.

Table 1: Experimental parameters

| Aspect | Quantity |
|---|---|
| Ion source | Hot filament driven volume type |
| Extraction | Triode extraction |
| Ion beam species | $He^+$, $H^+$, $H_2^+$, $H_3^+$ |
| Beam energy | 20 keV max |
| $He^+$ | 2.0 mA @ 10 keV |
| $H_3^+$ | ~ 95 % => 3.0 mA @ 10 keV |
| $H_2^+$ | ~ 91 % => 2.8 mA @ 10 keV |
| $H^+$ | ~ 45 % => 2.8 mA @ 10 keV |
| Toroidal magnetic field | On axis 0.6 T max |
| Major Radius | 1300 mm |
| Arc angle | 30 degree |
| Arc length | 680 mm |
| Toroid aperture | 200 mm |

For beam diagnostics an optical assembly was designed. It consisted of a phosphor screen mounted downstream of the sector magnet with repeller electrode. A digital camera was used to capture the image from screen. This fixed diagnostic assembly was installed downstream of the toroid.

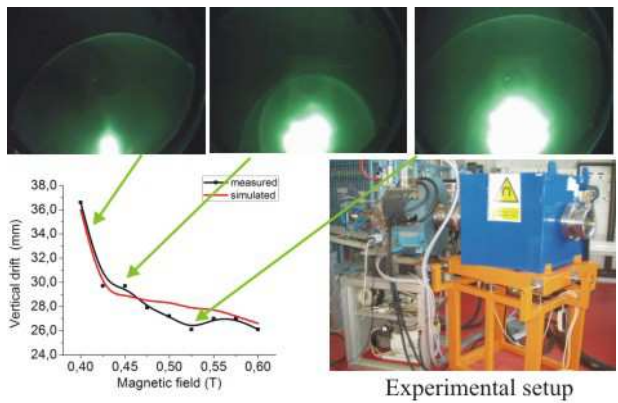

Figure 2: Measured vertical drifts at the end of toroid as a function of B-field.

The particle drifts were measured with this method as a function of magnetic field. Fig. 2 shows that the results are in good agreement with simulations. Results in detail are discussed in [3].

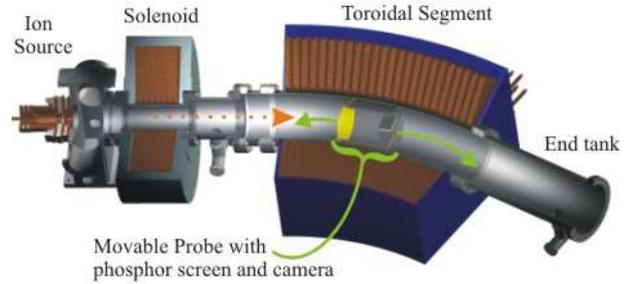

Figure 3: The setup showing ion source, solenoid, toroidal sector magnet and the movable optical probe.

An innovative "Pigging technique" was developed for beam diagnostics. The optical assembly was upgraded so as to manoeuvre the optical probe along the beam. The phosphor screen and the camera were installed on a movable cylinder (see Fig. 3). This assembly is able to sustain a magnetic field up to 0.6 T and high vacuum conditions. A repeller ring (± 1.2 kV) in front of the screen is used to get rid of secondary electrons produced due to beam losses. This gives an opportunity to detect the transversal beam profile along longitudinal axis.

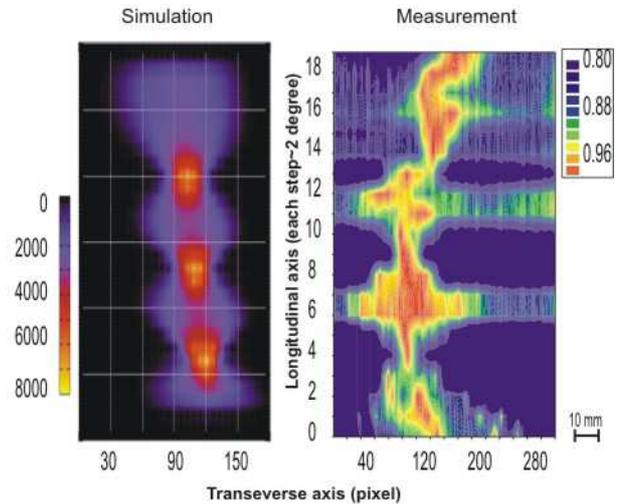

Figure 4: Transverse beam profile of the ion beam along the beam path, simulations and measurements.

Fig. 4 compares the simulation results with transverse profile measurement along a longitudinal axis. $He^+$-beam with energy of 6 keV was injected into a magnetic field of 0.6 T. As the beam is transported into the segment it shows periodic beam waists. The number of beam waists along the path can be calculated from beam energy and field strengths. Simulations show a very good agreement with measurements.

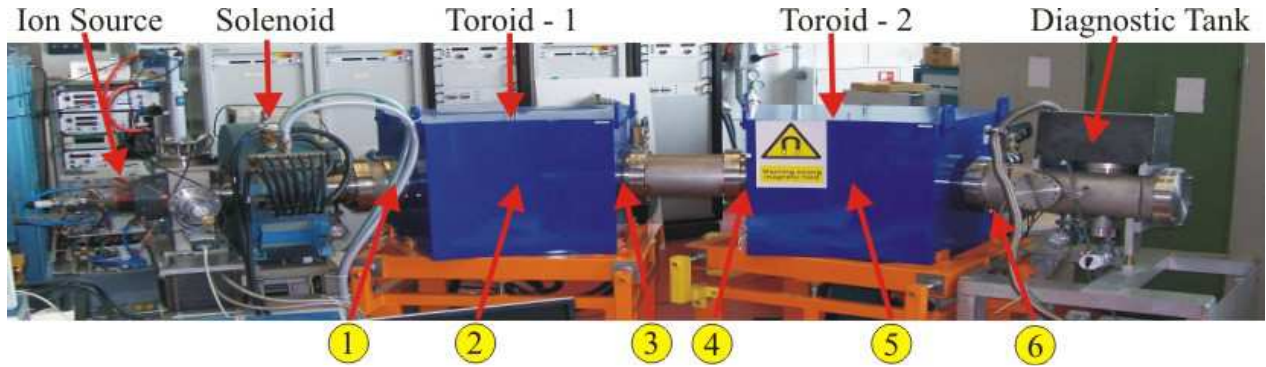

Figure 5: Experimental setup for beam transport experiments along two coupled toroidal segments.

*Beam Transport in coupled segments*

Two segments were coupled in the next stage (see Fig. 5). The magnetic field forms a ripple like structure.

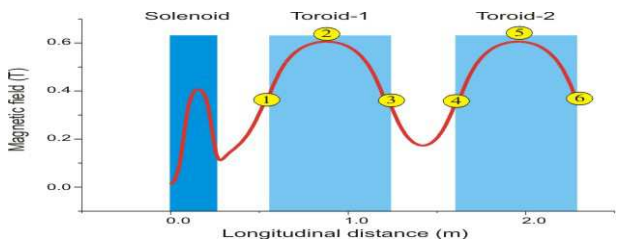

Figure 6: Magnetic field along longitudinal axis.

Fig. 6 shows the magnetic field strength along the longitudinal beam path. In the present setup, an intermediate distance of 400 mm is chosen between two segments as a test. This introduces a field drop of about 70%.

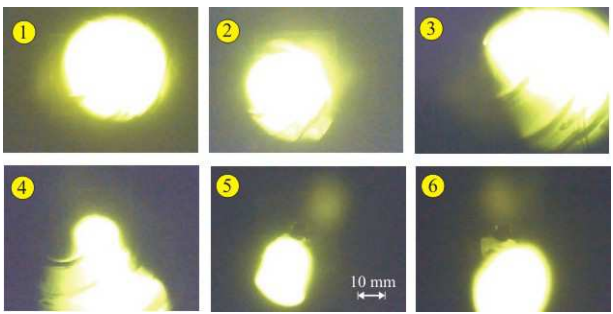

Figure 7: Measurement of proton beams at 9.70 keV energy transported through two coupled segments.

Fig. 7 shows a low energy proton beam (9.70 keV, 2.2 mA) detected at various positions along the beam path. The positions 3 and 4 clearly show beam loss due to ripple structure. The vertically drifted position along the beam path (1-6) can be seen is due to the curvature drift.

Fig. 8 A) shows the transmission (detected by optical screen intensity) as a function of longitudinal position of the probe which indicates heavy losses.

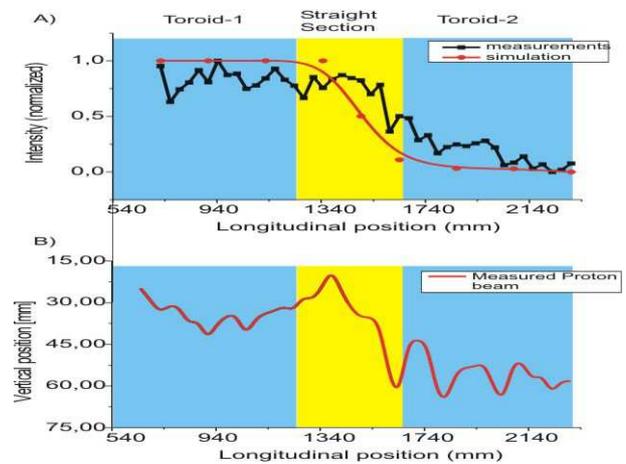

Figure 8: A) Intensity simulated and measured, B) Measured vertical drift plotted along longitudinal path.

Fig. 8 B) shows the vertically drifted position of an ion beam. The measurement is in reasonable agreement with the analytical value from Eq. (1) resulting in 32 mm.

## OUTLOOK

In this paper we have described the experimental activities under the project Stellarator Type Magnetostatic Storage Ring. The ion beam transport through two magnetically coupled segments has been successfully investigated. The effect of the distance between two magnets on beam transport will be studied in detail. The best suited case will be used for injection experiments, the working principle of which is described in [3].